\documentclass[aps,onecolumn,tightenlines,amsmath,amssymb,11pt,superscriptaddress,nofootinbib]{revtex4}

\usepackage{graphicx}
\usepackage{amsmath,amssymb,amsfonts,amsthm,stmaryrd,mathtools,bm,physics}
\usepackage{soul}
\usepackage{xcolor}
\usepackage{tikz}
\allowdisplaybreaks[1]
\usepackage[bookmarks,linktocpage, colorlinks=true, plainpages = false, citecolor = treegreen,  linkcolor=darkblue, urlcolor = darkblue, filecolor = blue]{hyperref} 

\definecolor{tealgreen}{rgb}{0.0, 0.5, 1.0}
\definecolor{darkblue}{rgb}{0., 0.4, 0.8}
\definecolor{cadmiumred}{rgb}{1., 0., 0.22}
\definecolor{treegreen}{rgb}{0., 0.7, 0.3}
\definecolor{emerald}{rgb}{0.31, 0.78, 0.47}
\definecolor{purple}{rgb}{1,0,1}
\definecolor{lime}{HTML}{A6CE39} 

\usepackage{pifont}

\def\be#1\ee{\begin{align}#1\end{align}}

\def\ba{\begin{eqnarray}}
\def\ea{\end{eqnarray}}
\def\nn{\nonumber}

\usepackage[normalem]{ulem}

\begin{document}

\title{Timelike convergence condition in regular black-hole spacetimes \\
	with (anti-)de~Sitter core}

\author{Johanna Borissova}
\email{jborissova@pitp.ca}
\affiliation{Perimeter Institute for Theoretical Physics, 31 Caroline Street North, Waterloo, ON, N2L 2Y5, Canada}
\affiliation{Department of Physics and Astronomy, University of Waterloo, 200 University Avenue West, Waterloo, ON, N2L 3G1, Canada}
\author{Stefano Liberati}
\email{liberati@sissa.it}
\affiliation{SISSA, Via Bonomea 265, 34136 Trieste, Italy}
\affiliation{INFN, Sez.~Trieste, Via Valerio 2, 34127 Trieste, Italy}
\affiliation{IFPU--Institute for Fundamental Physics of the Universe, 
Via Beirut 2, 34014 Trieste, Italy}
\author{Matt Visser}
\email{matt.visser@sms.vuw.ac.nz}
\affiliation{School of Mathematics and Statistics, Victoria University of Wellington, PO Box 600, Wellington 6140, New Zealand}

\begin{abstract}
{\sc Abstract:} The Hawking--Penrose (1970) singularity theorem weakens the causality assumption of global hyperbolicity used in the Penrose (1965) singularity theorem, at the expense of invoking the stronger timelike (instead of null) convergence condition (TCC). We analyze the TCC for a large class of dynamical spherically symmetric spacetimes, and show that it decomposes into three independent conditions on the Misner--Sharp mass function. For stationary black holes only two of these
are non-trivial. One of these conditions is already implied by the null convergence condition (NCC), whereas the other one depends explicitly on the TCC and constrains the sign of the second derivative of the mass function. We show that generic asymptotically flat regular black holes with a smooth de~Sitter core locally violate this new TCC-induced condition near the core, even if they globally satisfy the other condition imposed by the NCC. Therefore, it is the violation of the TCC which ensures that regular de~Sitter core black holes circumvent the Hawking--Penrose theorem. By contrast, we show that asymptotically flat regular black holes with an anti-de~Sitter core locally satisfy the new TCC-induced condition near the core, but necessarily violate it at some finite distance away from {it}. 
As concrete examples for both types of spacetimes, we consider TCC violations in the Bardeen black-hole spacetime with a de~Sitter core, and {in} a modified Bardeen black-hole spacetime with an anti-de~Sitter core. 
\\


\noindent
{\sc Date:} \today

\end{abstract}

\maketitle

\newpage 

\tableofcontents

\bigskip

\bigskip
\section{Introduction}\label{Sec:Introduction}

The classical stationary black holes of general relativity exhibit spacetime singularities. The presence of the singularity in the Schwarzschild black-hole spacetime can be deduced from the Penrose (1965) theorem~\cite{Penrose:1964wq}, see also references~\cite{Senovilla:2014gza,Hawking:1973uf}. 
This particular theorem is, however, inapplicable to the charged Reissner--Nordstr\"om and rotating Kerr black holes, due to the presence of a Cauchy horizon in these spacetimes. The presence of the Cauchy horizon violates global hyperbolicity, which is one of the key assumptions of the Penrose theorem. At the same time, this is the sole assumption which ensures that the standard stationary regular black holes with a smooth de~Sitter (dS) core, such as the Bardeen, Dymnikova, and Hayward black holes~\cite{Bardeen:1968bh,Dymnikova:1992ux,Hayward:2005gi}, which contain both an outer and an inner (Cauchy) horizon, circumvent the Penrose theorem. Indeed, these types of regular black holes globally satisfy the null convergence condition (NCC), which is the second key component of the Penrose theorem,  see e.g.~\cite{Borissova:2025msp}. On the other hand, as we will show, the NCC is violated in stationary regular black holes with a smooth anti-de~Sitter (adS) core, such as the adS-Bardeen black hole~\cite{Arrechea:2025xxx}. 
Taken together, the previous observations call for a singularity theorem which weakens the causality assumption of global hyperbolicity, at the expense of imposing a stronger convergence condition than the NCC. The Hawking--Penrose (1970) singularity theorem~\cite{Hawking:1970zqf}, see also references~\cite{Senovilla:2014gza,Hawking:1973uf}, does exactly this.\\

Specifically, the Hawking--Penrose theorem establishes the timelike or null geodesic incompleteness of a spacetime $(\mathcal{M},g)$ under the following conditions:
\begin{itemize}
\item[i)] Timelike and null convergence condition (TCC and NCC):
$R_{\mu\nu}v^\mu v^\nu \geq 0$ for all causal (i.e.~timelike or null) vectors $v^\mu$.
\item[ii)]Generic condition: Every causal geodesic with tangent vector $v^\mu$ contains a point at which $v_{[\mu}R_{\nu]\rho\sigma[\kappa}v_{\tau]}v^\rho v^\sigma \neq 0$.~\footnote{The generic condition for timelike curves is equivalent to the statement that the tidal tensor, defined by  $X_{\mu\nu} = R_{\mu\rho\nu\sigma} v^\rho v^\sigma$, is non-zero somewhere along the timelike curve.
More subtly, the generic condition for null curves is equivalent to the statement that the tidal tensor, now defined by  $X_{\mu\nu} = R_{\mu\rho\nu\sigma} k^\rho k^\sigma$ for any null vector $k^\mu$, is somewhere along the null curve \emph{not} of the ``trivial'' form 
$X_{\mu\nu} = k_\mu z_\nu + z_\mu k_\nu$, where $z^\mu$ is an arbitrary vector $4$-orthogonal to $k^\mu$.}
\item[iii)] Chronology condition: $\mathcal{M}$ has no closed timelike curves. 
\item[iv)]{$\mathcal{M}$ contains a closed trapped surface $\mathcal{T}$.}
\end{itemize}
It is important to note that the TCC for all timelike vectors already implies the NCC for all null vectors by a limiting argument. We will illustrate and make use of this fact explicitly in a spherically symmetric setup.

The assumptions of the above Hawking--Penrose theorem are satisfied for the stationary Reissner--Nordstr\"om black-hole spacetime, which is therefore geodesically incomplete. Unfortunately, the causality condition above is still too strong to allow the theorem to be applicable to the Kerr spacetime.  

Nevertheless, the advantage of the above Hawking--Penrose theorem is that the TCC is explicitly violated in stationary regular black holes with an inner Cauchy horizon and smooth dS cores,  examples for which are the Bardeen~\cite{Bardeen:1968bh}, Dymnikova~\cite{Dymnikova:1992ux} and Hayward~\cite{Hayward:2005gi} black holes. These spacetimes satisfy the NCC globally, see~e.g.~\cite{Borissova:2025msp}, but violate the additional assumption which is implied by the TCC beyond the NCC, as we will analyse in detail. By contrast, stationary regular black holes with an inner Cauchy horizon and an adS core, such as the modified Bardeen spacetime proposed in~\cite{Arrechea:2025xxx}, already violate the NCC, as part of the TCC, globally. Moreover, they violate the additional assumption imposed by the TCC in a finite region away from the core, as we will see.
Thus, altogether regular black holes with dS cores differ profoundly from regular black holes with adS cores in regards to the TCC and the NCC, as a limit of the TCC. Our main aim in this paper is to establish and contrast convergence conditions for both classes of regular black holes. Such considerations are key towards solidifying non-singular paradigms for black-hole physics~\cite{Carballo-Rubio:2025fnc,Buoninfante:2024oxl}.

This article is structured as follows. In Section~\ref{Sec:TCCGeneral} we derive the general conditions implied by the TCC for a class of spherically symmetric spacetimes. Section~\ref{Sec:TCCStationaryRBH} we analyse these conditions in detail for stationary regular black holes with a smooth dS or adS core. We close with a discussion in Section~\ref{Sec:Discussion}.

\section{TCC in spherically symmetric spacetimes}\label{Sec:TCCGeneral}

Let us consider a subclass of spherically symmetric spacetimes with line element in 
ingoing Eddington--Finkelstein coordinates
\be\label{eq:MetricImplodingSphericalSymmetry}
\dd{s^2} = -f(r,v)\dd{v}^2 + 2  \dd{v}\dd{r} + r^2   \dd{\Omega}^2 \qquad \text{where} \qquad f(r,v) = 1- \frac{2m(r,v)}{r}\,.
\ee
Here $\dd{\Omega}^2$ is the area element on the unit 2-sphere. This class of metrics is general enough to describe standard regular black holes with a smooth dS or adS core, see e.g.~\cite{Bardeen:1968bh, Dymnikova:1992ux,Hayward:2005gi,Arrechea:2025xxx}.

The timelike convergence condition (TCC) requires that
\be\label{eq:TCC}
R_{\mu\nu} \,v^\mu v^\nu \geq 0 \quad\text{for {\it any} timelike vector} \,\,\, v^\mu\,.
\ee
In spherical symmetry, ordering the coordinates as $\{v,r,\theta,\phi\}$,  any vector can without loss of generality be expressed in the form $v^\mu = (\alpha,\beta,\gamma,0)$. Imposing the normalization condition $v_\mu v^\mu = -\eta$ for some constant $\eta > 0$, and rescaling $\alpha \to 1$, we obtain
\be\label{eq:TimelikeVector}
v^\mu = \qty(1,-\frac{1}{2}\qty(\eta - f + \gamma^2 r^2),\gamma,0)\,.
\ee
The vector $v^\mu$ becomes a null vector in the limit $\eta \to 0$, which will allow us later to analyze the NCC as a limit of the TCC. Contracting $v^\mu$ with the Ricci tensor leads to
\be\label{eq:Rvv}
R_{\mu\nu} v^\mu v^\nu = \frac{1}{2r}\qty[\eta\qty(2 f' + r f'')  + \gamma^2\qty(-2 f r + f'' r^3 + 2 r) - 2 \dot{f}]\,,
\ee
where primes and dots denote partial derivatives with respect to $r$ and $v$. 

Generically, in the expression~\eqref{eq:Rvv} the constant $\eta\geq0$ and function $\gamma$ are arbitrary. Therefore, the TCC as stated in equation~\eqref{eq:TCC} can only be satisfied if all three terms in the square brackets are non-negative independently,
\ba
\frac{\eta}{2r}\qty(2 f' + r f'') &=& - \frac{\eta m''}{r} \geq 0 \,,\label{eq:TCCCondition1}\\
\frac{\gamma^2}{2 r}\qty(-2 f r + f'' r^3 + 2 r) &=& \gamma^2 \qty(2 m' - r m'') \geq 0\,,\label{eq:TCCCondition2}\\
-\frac{\dot{f}}{r} &=& \frac{2\dot{m}}{r^2} \geq 0\,.\label{eq:TCCCondition3}
\ea
Altogether, the conditions imposed by the TCC on the mass function $m(r,v)$ are
\be\label{eq:TCC3Conditions}
- m'' \geq 0 \,\,\, \quad \quad \text{and} \quad \quad \,\,\, 2 m' - r m'' \geq 0 \,\,\, \quad \quad \text{and} \quad\quad  \,\,\, \dot{m} \geq 0\,.
\ee

The first condition applies only when $\eta > 0$, but not in the limit $\eta \to 0$ when $v^\mu$ becomes a null vector. Thus, this is the only condition that is enforced by the TCC {\it in addition} to the second and third conditions, which are already enforced by the NCC. For a stationary geometry the third condition is trivially satisfied.  In Section~\ref{Sec:TCCStationaryRBH} we will analyze the first two inequalities in~\eqref{eq:TCC3Conditions} in detail for stationary regular black holes with a smooth dS or adS core. Before closing this section, we will provide  two examples of stationary singular black-hole spacetimes in which these conditions, as part of the Hawking--Penrose singularity theorem, are satisfied globally.

\subsection{Reissner--Nordstr\"om black hole}

\begin{figure}[h!]
	\centering
	\hspace{1.8cm}
	\includegraphics[width=0.75\textwidth]{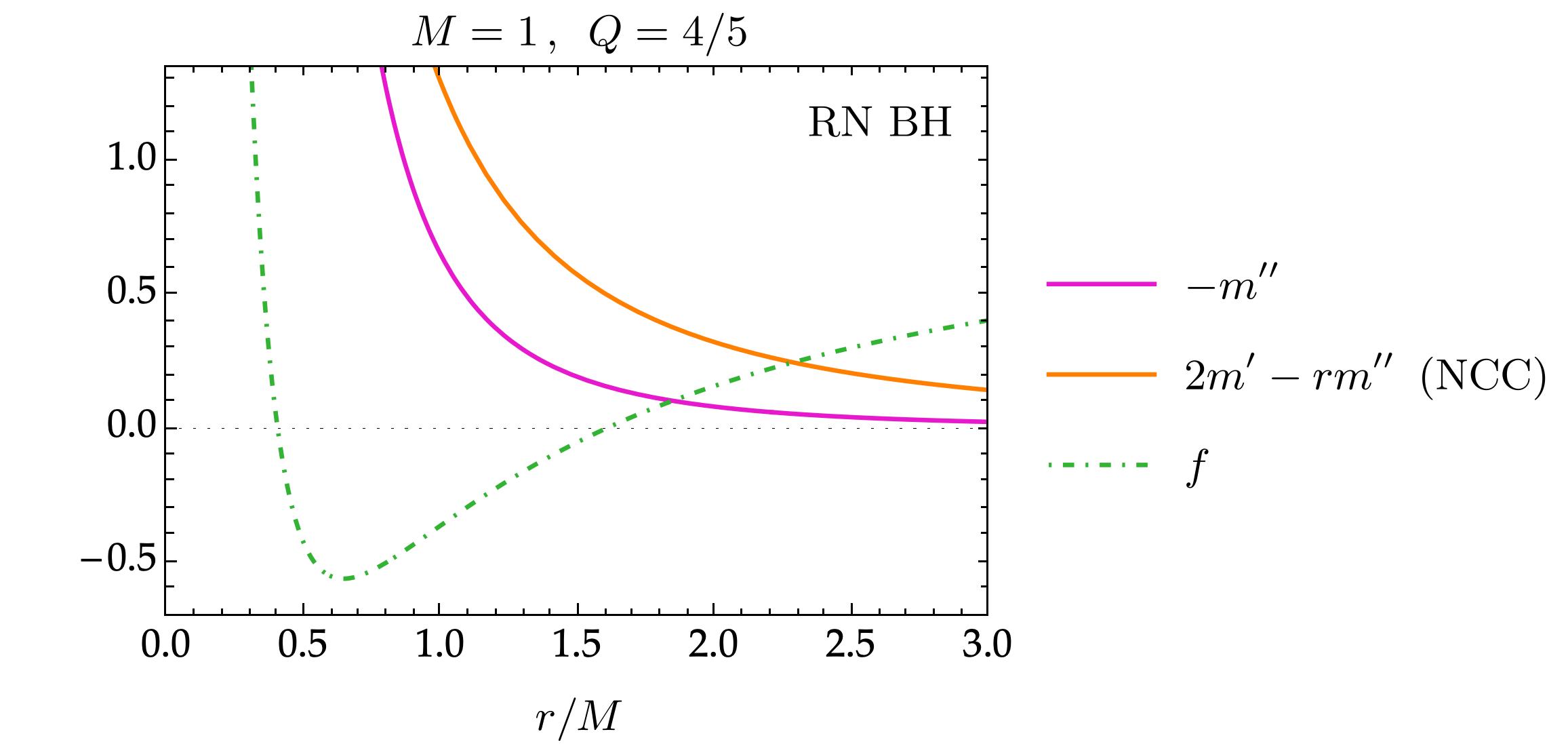}
	\caption{\label{Fig:RNBH} TCC expressions $-m''$  and $2 m' - r m''$, and lapse function $f$ for a Reissner--Nordstr\"om black hole, defined by the mass function~\eqref{eq:mBHIntegrable}. The mass and charge parameter are set to $M=1$ and $Q=4/5$. For this singular black-hole spacetime the TCC inequalities in equation~\eqref{eq:TCC3Conditions} are satisfied globally.}
\end{figure}

A familiar example for a singular black hole spacetime, in which the previous conditions are satisfied, is the Reissner--Nordstr\"om black hole with ADM mass parameter $M$ and charge $Q$, for which
\ba
m_{\text{RN}}(r) &=& m - \frac{Q^2}{2 r}\,,\label{eq:mRH}\\
-m_{\text{RN}}''(r) &=& \frac{Q^2}{r^3} >0\,\,\,\qquad \text{and} \,\,\, \qquad 
2 m_{\text{RN}}'(r) -r m_{\text{RN}}''(r) = \frac{2 Q^2}{r^2} > 0\,.
\ea
The expressions in the last line and the lapse function are shown in Fig.~\ref{Fig:RNBH}. The TCC is satisfied everywhere. \\

\subsection{Black hole with integrable singularity}

\begin{figure}[h!]
	\centering
	\hspace{1.8cm}
	\includegraphics[width=0.75\textwidth]{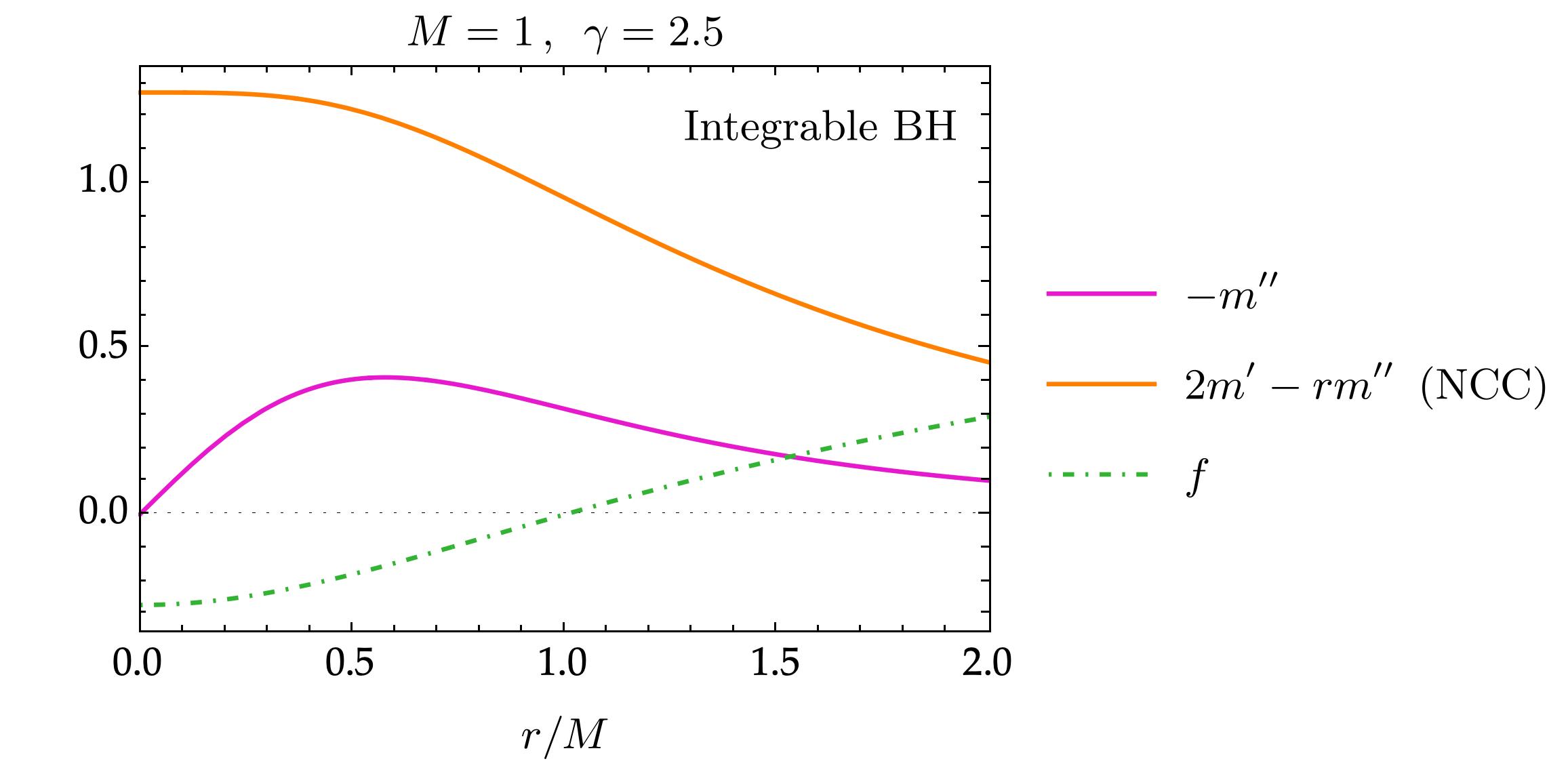}
	\caption{\label{Fig:IntegrableBH} TCC expressions $-m''$  and $2 m' - r m''$, and lapse function $f$ for a black hole with an integrable singularity, defined by the mass function~\eqref{eq:mBHIntegrable}. The parameter  $\gamma$ is set to $\gamma = 2.5$. For this singular black-hole spacetime the TCC inequalities in equation~\eqref{eq:TCC3Conditions} are satisfied globally.}
\end{figure}

An example of a less well-known class of singular black-hole spacetimes are black holes with an integrable singularity~\cite{Lukash:2011hd,Arrechea:2025fkk}, such as the spacetime proposed in~\cite{Casadio:2023iqt}. 
For this black-hole spacetime, the mass  function and quantities relevant for the TCC  are given by
 \ba
 m_{\text{BH}}(r) &=& \frac{2 M}{\pi} \arctan(\frac{\gamma r}{M})\label{eq:mBHIntegrable}\,,\\
  -m_{\text{BH}}''(r) &=& \frac{4 M^2 }{\gamma \pi}\frac{r}{\qty( r^2 + \frac{M^2}{\gamma^2})^2} >0\,,\\
   2 m_{\text{BH}}'(r) -r m_{\text{BH}}''(r) &=& \frac{4  M^2}{\gamma \pi } \frac{2  r^2 + \frac{M^2}{\gamma^2}}{\qty( r^2 + \frac{M^2}{\gamma^2})^2} >0\,.\quad \quad \quad \quad \quad \quad 
 \ea
The expressions in the last two lines and the lapse function are shown in Fig.~\ref{Fig:IntegrableBH}. The TCC holds everywhere, albeit the first condition in~\eqref{eq:TCC3Conditions} being only marginally satisfied at the (integrable) singularity. 

\section{TCC violation in regular black-hole spacetimes with (anti)-de~Sitter core}
\label{Sec:TCCStationaryRBH}

In this section we restrict the analysis to a stationary geometry $(\dot{m} = 0)$ and consider the first two conditions in~\eqref{eq:TCC3Conditions} on the Misner--Sharp quasi-local mass $m(r)$ for regular black-hole spacetimes with a smooth de~Sitter (dS) or anti-de~Sitter (adS) core. The mass function for these types of regular black holes can be expanded at $r=0$ as a power series
\ba\label{eq:MExpansion}
m(r) &=& \sum_{n=0}^\infty m_n \,r^n =m_0 + m_1\, r + m_2\, r^2 + m_3\, r^3 + m_4\,r^4 +\mathcal{O}\qty(r^5)\,,\\
m'(r) &=& \sum_{n=0}^\infty (n+1) m_{n+1} \,r^{n} = m_1  + 2 m_2 \, r + 3 m_3 \,r^2 + 4 m_4 \,r^3 + \mathcal{O}\qty(r^4)\,,\\
m''(r) &=& \sum_{n=0}^\infty (n+2) (n+1) m_{n+2} \,r^{n} = 2 m_2 + 6 m_3\, r + 12 m_4 \,r^2 + \mathcal{O}\qty(r^3)
\ea
with the condition $m_0 = m_1 = m_2 \equiv 0$ for regularity, see e.g.~\cite{Carballo-Rubio:2019fnb}, and leading coefficient $m_3 \neq 0$. The physical interpretation of the parameter $m_3$ becomes clear by expressing the metric function $f$ to leading orders at small $r$ as
\be
f(r) = 1 - \frac{m(r)}{r} = 1 - m_3 \, r^2 + \mathcal{O}\qty(r^3)\,.
\ee
Thus,  $m_3 \longleftrightarrow \Lambda$ plays the role of a cosmological constant whose sign determines whether the spacetime at the core is dS $(\Lambda >0)$ or adS $(\Lambda <0)$. This defines the notion of a regular black hole with a dS or adS core.\\

For this class of regular black holes the first two conditions imposed by the TCC in~\eqref{eq:TCC3Conditions} can be expressed locally (near $r=0$) to leading order as
\ba 
-m''(r) &\geq & 0 \,\,\, \quad \Longleftrightarrow \quad  \,\,\, -m_3 > 0 \label{eq:TCC1m3}\,,\\
2 m'(r) - r m''(r)  &\geq & 0 \,\,\, \quad \Longleftrightarrow \quad  \,\,\, -m_I > 0 \label{eq:TCC2mI}\,.
\ea
Here $m_I$ is the first non-zero coefficient in the expansion of $m(r)$ at subleading order. We remind the reader that the second condition~\eqref{eq:TCC2mI} is already imposed by the NCC, whereas the first condition~\eqref{eq:TCC1m3} is a condition arising only as a result of considering the TCC, rather than the NCC.

\subsection{Regular black holes with de~Sitter core}\label{SecSub:RBHdS}

The mass function for regular black holes with a dS core can be expanded at $r=0$ as
\be\label{eq:mBHdS}
m_{\text{BHdS}}(r) = m_3 \, r^3 + \mathcal{O}\qty(r^4)\,,\quad m_3   > 0\,.
\ee
The first condition~\eqref{eq:TCC1m3} imposed by the TCC in this case is violated near $r=0$, since $-m_3  < 0$. 

Moreover, according to the generic expansion of the mass function~\eqref{eq:MExpansion}, we see that $m_{\text{BHdS}}'(0)\equiv0$ but $m_{\text{BHdS}}'(r)>0$ in a neighborhood of $r=0$. Assuming that the spacetime is asymptotically flat, with positive Arnowitt--Deser--Misner (ADM) mass parameter $M$, we also know that $m_{\text{BHdS}}(r)=M>0$ and $m_{\text{BHdS}}'(r)\equiv 0$ for $r\to \infty$. Thereby, we can conclude that $m_{\text{BHdS}}'$ must have at least one local maximum somewhere in the interval $(0,\infty)$. Let $r=r_{\text{TCC}}$ be the minimum of all possible values for $r$ corresponding to a local maximum of $m_{\text{BHdS}}'$. Then $m_{\text{BHdS}}''$ will change sign at $r=r_{\text{TCC}}$, which marks the boundary of the TCC violating region $r\in (0,r_{\text{TCC}})$ near the core where  $m_{\text{BHdS}}''(r) > 0$. We shall give an explicit example of such behaviour below. Altogether we can conclude:\\

{\it Asymptotically flat regular black holes with a de~Sitter core must violate the first TCC condition}, $-m_{\text{BHdS}}''\geq 0$, {\it in~\eqref{eq:TCC3Conditions} in some finite region around the core.}\\

Whether the second condition~\eqref{eq:TCC2mI}, already imposed by the NCC, is also violated, depends in principle on the specifics of mass function. However, for the well-known dS core regular black holes such as the Bardeen, Hayward or Dymnikova black holes~\cite{Bardeen:1968bh,Hayward:2005gi,Dymnikova:1992ux} this condition must be satisfied, given that the NCC is respected in these spacetimes globally, see e.g.~\cite{Borissova:2025msp}. For concreteness, below we provide an example which will also allow us later to exemplify the stark contrast between dS core and  adS core regular black holes with regard to the separate conditions~\eqref{eq:TCC3Conditions} imposed by the TCC.

\subsubsection{Example: Bardeen black hole with de~Sitter core}\label{SecSubSub:BardeendS}

\begin{figure}[h!]
	\centering
	\hspace{1.8cm}
	\includegraphics[width=0.75\textwidth]{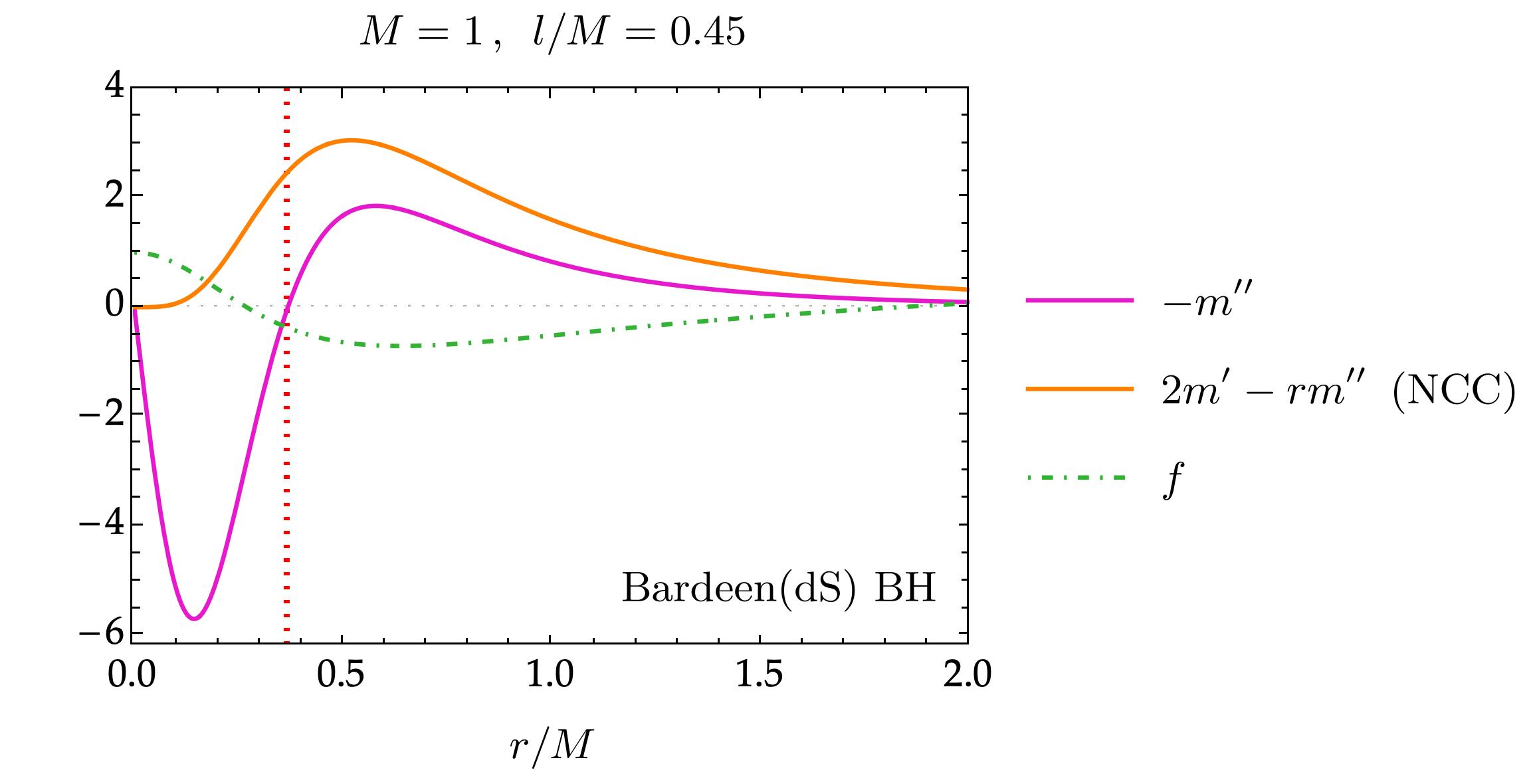}
	\caption{\label{Fig:BardeendS} TCC expressions $-m''$  and $2 m' - r m''$ for a Bardeen black hole with a dS core, defined by the mass function~\eqref{eq:MassFunctionBardeendS}. The regularization parameter is set to  $l/M=0.45$. While the NCC, given by the second condition in~\eqref{eq:TCC3Conditions}, is satisfied globally, the first condition in~\eqref{eq:TCC3Conditions} which is imposed additionally by the TCC, is violated near the core. The boundary of the TCC violating region is marked by the red dotted line at $r_{\text{TCC}} / M 
		\approx 0.367$.}
\end{figure}

As a concrete example, we consider the regular Bardeen black hole with a dS core~\cite{Bardeen:1968bh} with  mass function parametrized by the ADM mass parameter $M>0$ and a regularization length parameter $l>0$. For the Bardeen spacetime,
\ba
m_{\text{BdS}}(r) &=&  \frac{M r^3}{\qty(r^2 + l^2)^{3/2}} = + \frac{M}{l^3}r^3 - \frac{3 M}{2l^5} r^5 +\mathcal{O}\qty(r^7)  \label{eq:MassFunctionBardeendS}\,, \quad \quad \quad \quad \\
-m_{\text{BdS}}''(r) &=& \frac{3 M l^2  r}{\qty(l^2 r^2 + l^4)^{7/2}} \qty( 3 r^2 - 2 l^2 ) \label{eq:TCCExtraBardeen}\,,\\
2 m_{\text{BdS}}'(r) - r m_{\text{BdS}}''(r) &=& \frac{14 M l^2 r^4}{\qty(r^2 + l^2)^{7/2}} > 0\label{eq:NCCBardeen} \,.
\ea
From~\eqref{eq:MassFunctionBardeendS} we see on the one hand that $-m_I = -m_5 = 3 M/ \qty(2 l^5) > 0$. Thus, the NCC is indeed satisfied locally near the core, but also globally, as the expression in~\eqref{eq:NCCBardeen} shows. On the other hand, according to~\eqref{eq:MassFunctionBardeendS} it is in particular $-m_3 = -M/l^3 < 0$. Therefore, the ``new'' part of the TCC, which is not already imposed by the NCC, is violated. The boundary of the region of TCC violation near the core is defined by the root of~\eqref{eq:TCCExtraBardeen}, which is given by $r =\sqrt{2/3}\, l$. Fig.~\ref{Fig:BardeendS} shows the expressions in the last two lines above, as well as the lapse function, for the Bardeen black hole for a concrete choice of parameter $l/M=0.45$. In this example, the boundary of the TCC violating region defined by $m_{\text{BdS}}''(r_{\text{TCC}})=0$  lies within the trapped region $(r_-,r_+)$, defined by the zeros $r_\pm$ of the lapse function $f$. However, this does not need to be the case. Increasing the length parameter $l$, we can achieve a configuration with two horizons, for which the boundary of the TCC violating region, defined by $m_{\text{BdS}}''(r_{\text{TCC}})=0$, lies to the interior of the inner horizon and therefore in the {locally} untrapped region.\footnote{We use the phrase ``locally untrapped'' to emphasize just how odd the physics inside the inner horizon is. In that region ``outgoing'' light rays are truly outgoing, though they eventually accumulate on the inside of the inner horizon, implying a notion of ``quasi-local trapping''. {Note, however, that the standard singularity theorems only make use of the ``local trapping" of light/matter, and are hence oblivious to such ``quasi-local trapping".}}

\begin{figure}[h!]
 	\centering
 	\hspace{1.cm}
 	\includegraphics[width=0.65\textwidth]{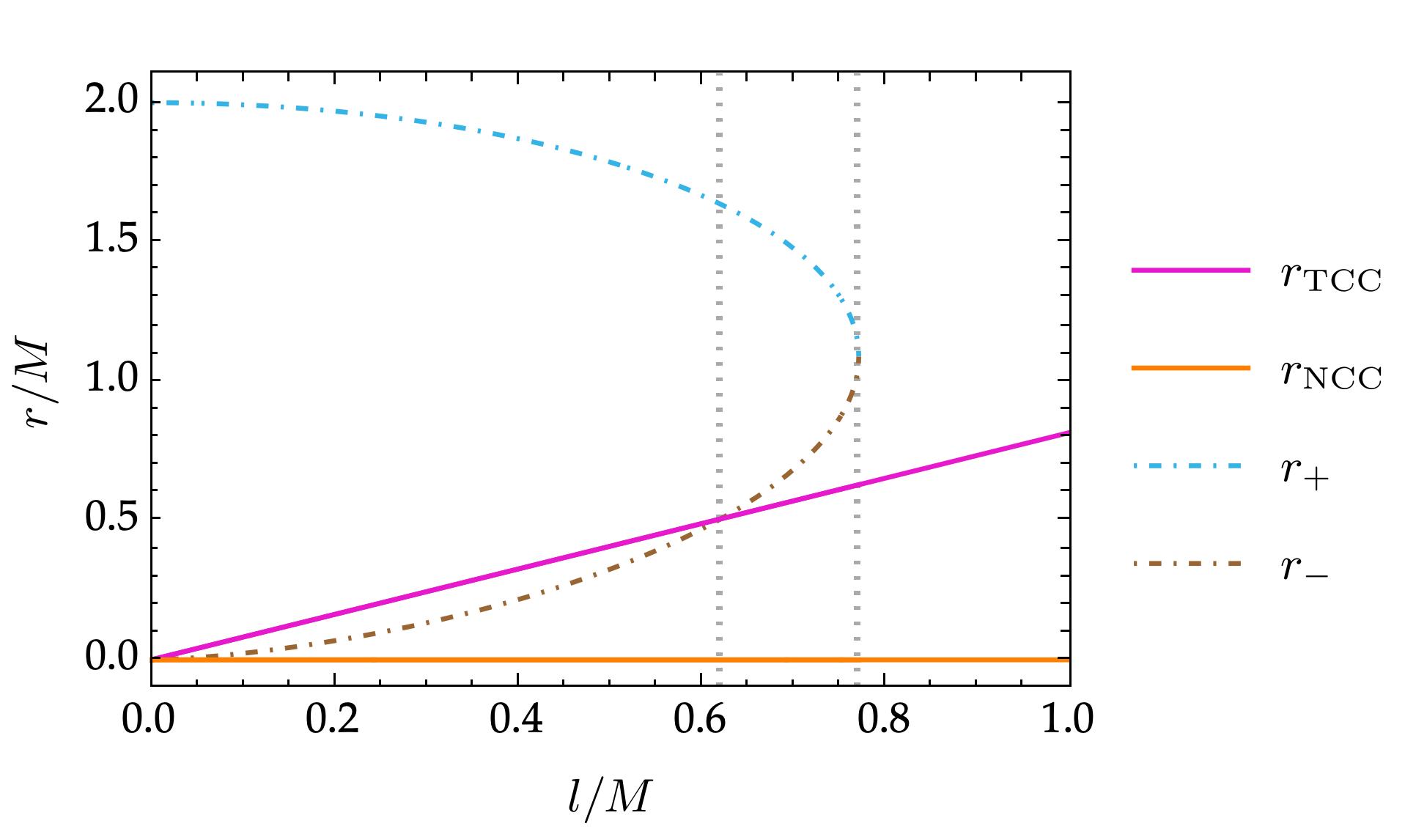}
 	\caption{\label{Fig:BardeenrTCCrh} Inner and outer horizons $r_\pm$, defined by $f(r_\pm)=0$, and boundary $r_{\text{TCC}}$ of the TCC violating region, defined by $m''(r_{\text{TCC}})=0$, as functions of the regularization parameter $l/M$ for a Bardeen black hole with a dS core. For small enough $l/M \in (0, 0.620)$, the boundary $r_{\text{TCC}}$ lies within the trapped region $(r_-,r_+)$, and coincides with the inner horizon  when $l/M \approx 0.620$. In turn, $r_{\text{TCC}}$ lies in the {locally} untrapped region {inside the inner horizon} for $l/M \in (0.620,0.770)$,  where $r_\pm/M  \approx 0.770$ marks the extremal configuration in which the two horizons degenerate to one. For larger values of $l/M$, the geometry does not describe black hole, but rather represents a horizonless compact object. For completeness, we also show the line $r_{\text{NCC}}\equiv 0$ where the NCC is saturated, i.e., where  $2m'(r_{\text{NCC}}) - r_{\text{NCC}}\, m''(r_{\text{NCC}}) =0$.}
 \end{figure}

More generally, the roots  $r_\pm$ of $f(r)$, which represent the locations of the inner and outer horizons for the Bardeen spacetime with mass function~\eqref{eq:MassFunctionBardeendS}, can be found analytically as functions of the regularization parameter $l$. Fig.~\ref{Fig:BardeenrTCCrh} shows $r_\pm$ and the boundary $r_{\text{TCC}}$ of the TCC violating region defined by $m_{\text{BdS}}''(r_{\text{TCC}})=0$ as functions of $l/M$. For small enough $l/M$, the boundary $r_{\text{TCC}}$ lies within the trapped region $(r_-,r_+)$, coincides with the inner horizon  when $l/M \approx 0.620$ and lies in the {locally} untrapped region {inside the inner horizon} for $l/M \in (0.620,0.770 )$, where $r_+/M = r_-/M \approx 0.770$ marks the extremal configuration in which the two horizons degenerate to one. For larger values of $l/M$, the geometry does not describe black hole {but rather a horizonless black-hole mimicker}. For completeness, Fig.~\ref{Fig:BardeenAdSrTCCrh} also shows the boundary $r_{\text{NCC}}\equiv 0$ where the NCC is saturated, i.e., where  $2m_{\text{BdS}}'(r_{\text{NCC}}) - r_{\text{NCC}} \, m_{\text{BdS}}''(r_{\text{NCC}}) =0$.\\

One may verify that qualitatively similar conclusions hold for the Dymnikova or Hayward black holes~\cite{Dymnikova:1992ux,Hayward:1994bu}. In particular, whereas these regular black-hole spacetimes satisfy the NCC globally,  near the core they violate the part of the TCC which is not imposed by the NCC. This TCC violation in turn ensures that the Hawking--Penrose singularity theorem is circumvented.

\subsection{Regular black holes with anti-de~Sitter core}\label{SecSub:RBHAdS}

The mass function for regular black holes with an adS core can be expanded at $r=0$ as
\be\label{eq:mBHAdS}
m_{\text{BHAdS}}(r) = m_3\; r^3 + \mathcal{O}\qty(r^4)\,,\quad m_3   < 0\,.
\ee
Therefore, according to~\eqref{eq:TCC1m3}, in this case the first condition imposed by the TCC is satisfied near $r=0$ since $-m_3 > 0$. However, this conclusion holds only locally near $r=0$, and a violation of the first inequality in~\eqref{eq:TCC3Conditions} can still occur at some finite distance away from the core. That this is indeed the case can be seen as follows.

 According to the generic expansion of the mass function~\eqref{eq:MExpansion}, $m_3 < 0$ implies that in a neighborhood of $r=0$ the first and second derivatives of the mass function satisfy $m_{\text{BHAdS}}'(r)<0$ and $m_{\text{BHAdS}}''(r)<0$. Assuming that the spacetime is asymptotically flat with positive ADM mass parameter $M$ we also know, in addition to $m_{\text{BHAdS}}(0)=0$, that $m_{\text{BHAdS}}(r) \to  M > 0$ for $r\to \infty$. Thereby we can conclude that $m_{\text{BHAdS}}'$ must change sign at least once in the interval $(0,\infty)$. Let $r=r_1$ be the minimum of all possible values for $r$ where a sign change of $m'$ occurs. It follows that $m_{\text{BHAdS}}'$ must have a local minimum at some value $r=r_{\text{TCC}-}$ where $ r_{\text{TCC}-} \in (0,r_1)$. In particular $m_{\text{BHAdS}}''$ will change sign at $r_{\text{TCC}-}$ and satisfy $m_{\text{BHAdS}}''(r)>0$ for $r \in (r_{\text{TCC}-},r_{\text{TCC}+})$ where $r_{\text{TCC}+}$ is defined to be the next local extremum (a local maximum) of $m'$ which lies to the exterior of $r_1$. That such a local maximum exists, follows by an analogous argument as the one given previously, namely one has  $m_{\text{BHAdS}}'(r_1)=0$ and $m_{\text{BHAdS}}'(r)=0$ for $r\to \infty$ but $m_{\text{BHAdS}}'(r)>0$ for $r>r_1$ infinitesimally close to $r_1$. Therefore in the region $(r_{\text{TCC}-},r_{\text{TCC}+})$ of the regular black-hole spacetime the first condition imposed by the TCC in~\eqref{eq:TCC3Conditions} will be violated. Altogether we can conclude:\\

{\it Asymptotically flat regular black holes with an anti-de~Sitter core necessarily violate the first TCC condition}, $-m_{\text{BHAdS}}'' \geq 0 $, {\it  in~\eqref{eq:TCC3Conditions} in a finite region away from the core.}

\subsubsection{Construction of regular black holes with anti-de~Sitter core}\label{SecSubSub:Construction}

As a concrete construction algorithm for asymptotically flat regular black holes with an adS core we {consider a proposal~\cite{Arrechea:2025xxx} starting from an asymptotically flat regular black hole with a dS core.}
Let  $m_{\text{BHdS}}(r)$ be its mass function and $l$ its characteristic regularization length parameter. We assume that  $m_{\text{BHdS}}(r) \to M$ for $l\to 0$ such that in this limit the spacetime asymptotes to the singular Schwarzschild spacetime. Additionally, $m_{\text{BHdS}}(r)\to M$ for $r\to \infty$ according to asymptotic flatness. Consider the product
\be
m(r) \equiv m_{\text{BHdS}}(r)\; \tau_n(r) \,\,\,\quad \text{where} \quad \,\,\,\tau_n(r) \equiv l^{2\qty(n-1)}\qty(\frac{r^2-l^2}{\qty(r^2 + l^2)^n})
\ee
for an \emph{a priori} arbitrary power $n$. By construction the function $\tau(r)$ is dimensionless. Expanding the function $\tau_n(r)$ at $r=0$ leads to
\ba
\tau_n(r) = -1 + \mathcal{O}\qty(r^2)\,.
\ea
This leading-order minus sign flips the sign of the mass function of the regular dS core black hole at $r=0$ and thereby leads to a regular black hole with an adS core. For the condition of asymptotic flatness we note that
\be
\tau_n(r) \simeq l^{2\qty(n-1)} + \mathcal{O}\qty(\frac{1}{r^2})
\ee
for large $r$. Thus, we must set $n=1$ in order to achieve $m(r)\to m_{\text{BHdS}}(r) \to M$ for $r\to\infty$.  Thereby we arrive a regular black hole with an adS core and mass function
\be
m_{\text{BHAdS}}(r) \equiv m_{\text{BHdS}}(r)\tau_1( r) =  m_{\text{BHdS}}(r)\qty(\frac{r^2-l^2}{r^2 + l^2}) \,.
\ee
Based on this mass function we can concretely analyse the first two inequalities in~\eqref{eq:TCC3Conditions}. Focusing on the first one, we compute
\ba
-m_{\text{BHAdS}}''(r) &=& -\frac{-4 m_{\text{BHdS}}(r)\qty(l^4 - 3l^2 r^2) - \qty(r^2 + l^2)\qty(8 l^2 r m_{\text{BHdS}}'(r) + \qty(r^4-l^4)m_{\text{BHdS}}''(r))}{\qty(r^2 + l^2)^3}\,\,\,\,\,\nn\\
&=& - \frac{4}{l^2} m_{\text{BHdS}}(\epsilon) + m_{\text{BHdS}}''(\epsilon) + \mathcal{O}\qty(r)\,,
\ea
where in the second line we have expanded around $r=0$ and used the 
notation $m_{\text{BHdS}}(\epsilon)$ and $m_{\text{BHdS}}''(\epsilon) $ for $\epsilon >0$ arbitrarily small to denote the respective series expansions of these functions around $r=0$ (this is to avoid writing $m_{\text{BHdS}}(0)$ and $m_{\text{BHdS}}''(0) $ as this evaluation would give zero). Making use of $m_{\text{BHdS}}(\epsilon) > 0$ and $m''_{\text{BHdS}}(\epsilon)<0$ for a regular dS core black hole, we can conclude $-m_{\text{BHAdS}}''(\epsilon) < 0$. Therefore, as expected, the first TCC condition is satisfied near $r=0$ for these classes of adS core regular black holes. Exactly where the TCC violating region occurs away from the core depends on the specifics of the mass function $m_{\text{BHdS}}(r)$ of the regular dS core black hole.

Next, we consider the second condition in~\eqref{eq:TCC3Conditions} associated with the NCC and to that end compute
\ba
2 m_{\text{BHAdS}}'(r) - r m_{\text{BHAdS}}''(r) &=& \frac{1}{\qty(r^2+l^2)^3}\Big(4 l^2 r\qty(5r^2+l^2)m_{\text{BHdS}}(r)\nn\\
&{}&\quad + \qty(r^2+l^2)\qty(2 \qty( r^4 - 4l^2r^2 -l^4)m_{\text{BHdS}}'(r) - r\qty(r^4-l^2)m_{\text{BHdS}}''(r))\Big)\nn\\
&=& - m'_{\text{BHdS}}(\epsilon)+\mathcal{O}\qty(r)\,.
\ea
Since $m'_{\text{BHdS}}(\epsilon)>0$ for regular dS core black holes we can conclude that the NCC, as part of the TCC, is violated locally near $r=0$ for these classes of adS core regular black holes.

\subsubsection{Example: Modified Bardeen black hole with anti-de~Sitter core}\label{SecSubSub:BardeenAdS}

\begin{figure}[h!]
	\centering
	\hspace{1.8cm}
	\includegraphics[width=.75\textwidth]{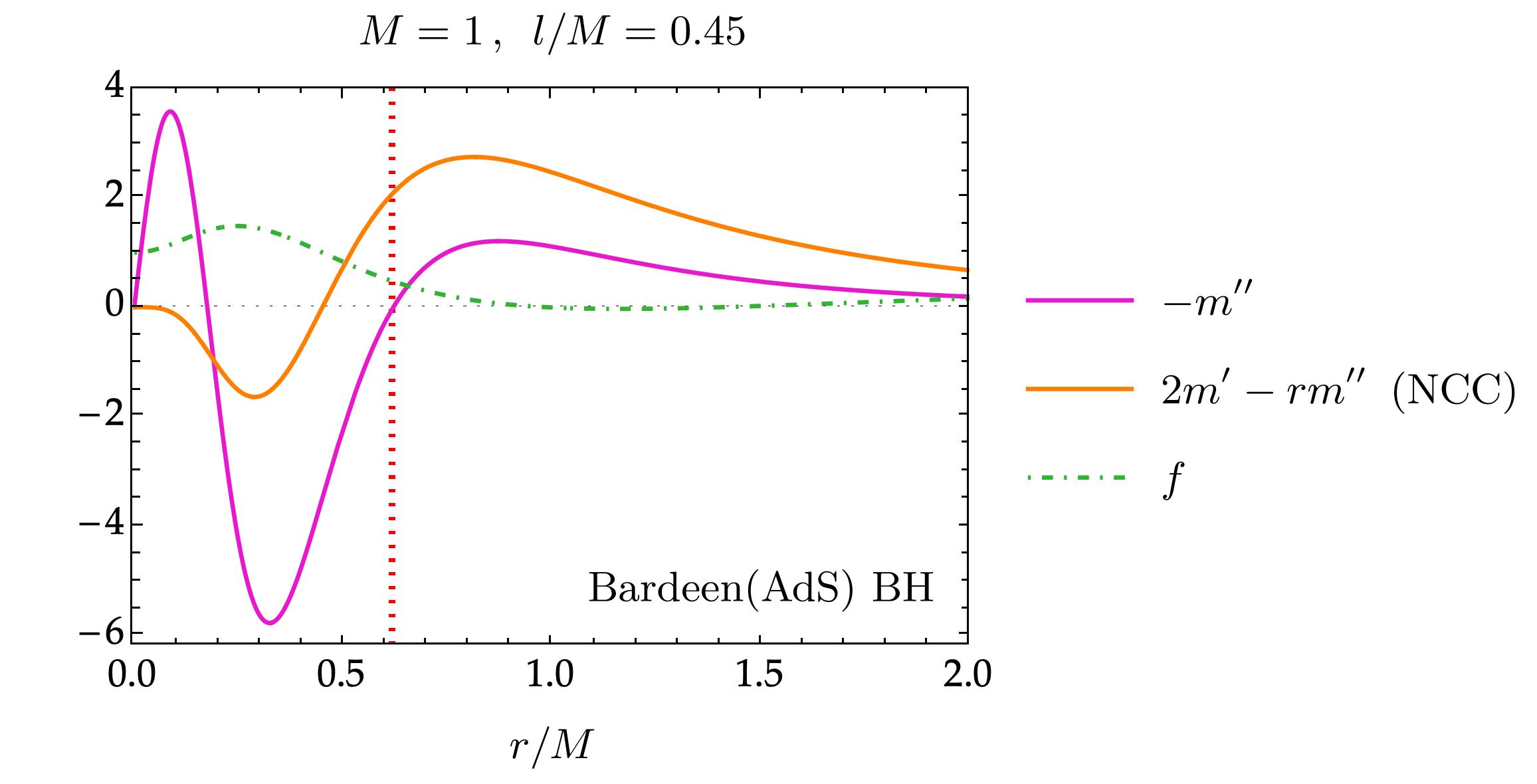}
	\caption{\label{Fig:BardeenAdS} TCC expressions $-m''$  and $2 m' - r m''$, and lapse function $f$ for a modified Bardeen black hole with an adS core, defined by the mass function~\eqref{eq:MassFunctionBardeenAdS}. The length parameter  is set to $l/M=0.45$. The NCC, given by the second condition in~\eqref{eq:TCC3Conditions}, is violated locally near the core, whereas the the first condition in~\eqref{eq:TCC3Conditions} which is imposed additionally by the TCC, is violated in a finite region away from the core. The outer boundary of the TCC violating region is marked by the red dotted line at $r_{\text{TCC+}} / M 
			 \approx 0.620$.}
\end{figure}

As a concrete example, we consider a modified Bardeen black hole with a smooth adS core~\cite{Arrechea:2025xxx}. 
The mass function is parametrized by the ADM mass parameter $M>0$ and a regularization length parameter $l>0$. For this modified Bardeen spacetime,
\ba
m_{\text{BAdS}}(r) &=& M \qty(\frac{r}{\sqrt{r^2 + l^2}})^3 \qty(\frac{r^2 - l^2}{r^2 + l^2}) = - \frac{M}{l^3}r^3 + \frac{7 M}{2l^5}r^5+ \mathcal{O}\qty(r^7)\label{eq:MassFunctionBardeenAdS} \,, \quad \quad \quad \\
-m_{\text{BAdS}}''(r) &=& \frac{M l^2 }{\qty(r^2 +l^2)^{9/2}}\qty(21  r^5 - 43 l^2 r^3 + 6 l^4 r) \label{eq:TCCExtraBardeenAdS}\,,\\
2 m_{\text{BAdS}}'(r) - r m_{\text{BAdS}}''(r) &=& \frac{35 M l^2 r^4}{\qty(r^2 + l^2)^{9/2}}\qty(r^2 - l^2)\label{eq:NCCBardeenAdS}\,.
\ea
It should be noted that this mass function is negative near $r=0$.
From~\eqref{eq:MassFunctionBardeenAdS} we see on the one hand $-m_I = -m_5 = - 7 M /\qty(2 l^5) <0$. Thus the NCC,  as part of the TCC, is violated locally near the core. The boundary $r_{\text{NCC}}$ of the region of NCC violation is defined by the root of~\eqref{eq:NCCBardeenAdS}, which is given by $r_{\text{NCC}}=l$. On the other hand, according to~\eqref{eq:MassFunctionBardeenAdS} it is in particular $-m_3 = M/l^3 > 0$. 
Therefore, that part of the TCC, which is not already imposed by the NCC, is locally satisfied near the core. However, this part of the TCC is violated in some finite region away from the core, which can be deduced from the  roots  of the expression in~\eqref{eq:TCCExtraBardeenAdS}, i.e., by solving $m_{\text{BAdS}}''(r_{\text{TCC}\pm}) = 0$. These roots are given by $r_{\text{TCC}\pm} = \sqrt{\frac{43 \pm \sqrt{1345}}{42}}\, l$. Fig.~\ref{Fig:BardeenAdS} shows the expressions in the last two lines above, as well as the lapse function for the modified Bardeen black hole for a concrete choice for the parameter $l/M$. In this example, the outer boundary of the region which violates the first TCC condition, $r_{\text{TCC}+}$, lies to the interior of the inner horizon defined by the smallest zero of $f$, and therefore in particular in the locally untrapped region surrounding the core. However, this does not need to be the case. Decreasing the length parameter $l$ we can achieve a configuration with two horizons, in which this boundary lies inside the trapped region. On the other hand, the boundary $r_{\text{NCC}}$ of the NCC violating region lies always in the locally untrapped region inside the inner horizon.

\begin{figure}[h!]
	\centering
	\hspace{1.cm}
	\includegraphics[width=0.65\textwidth]{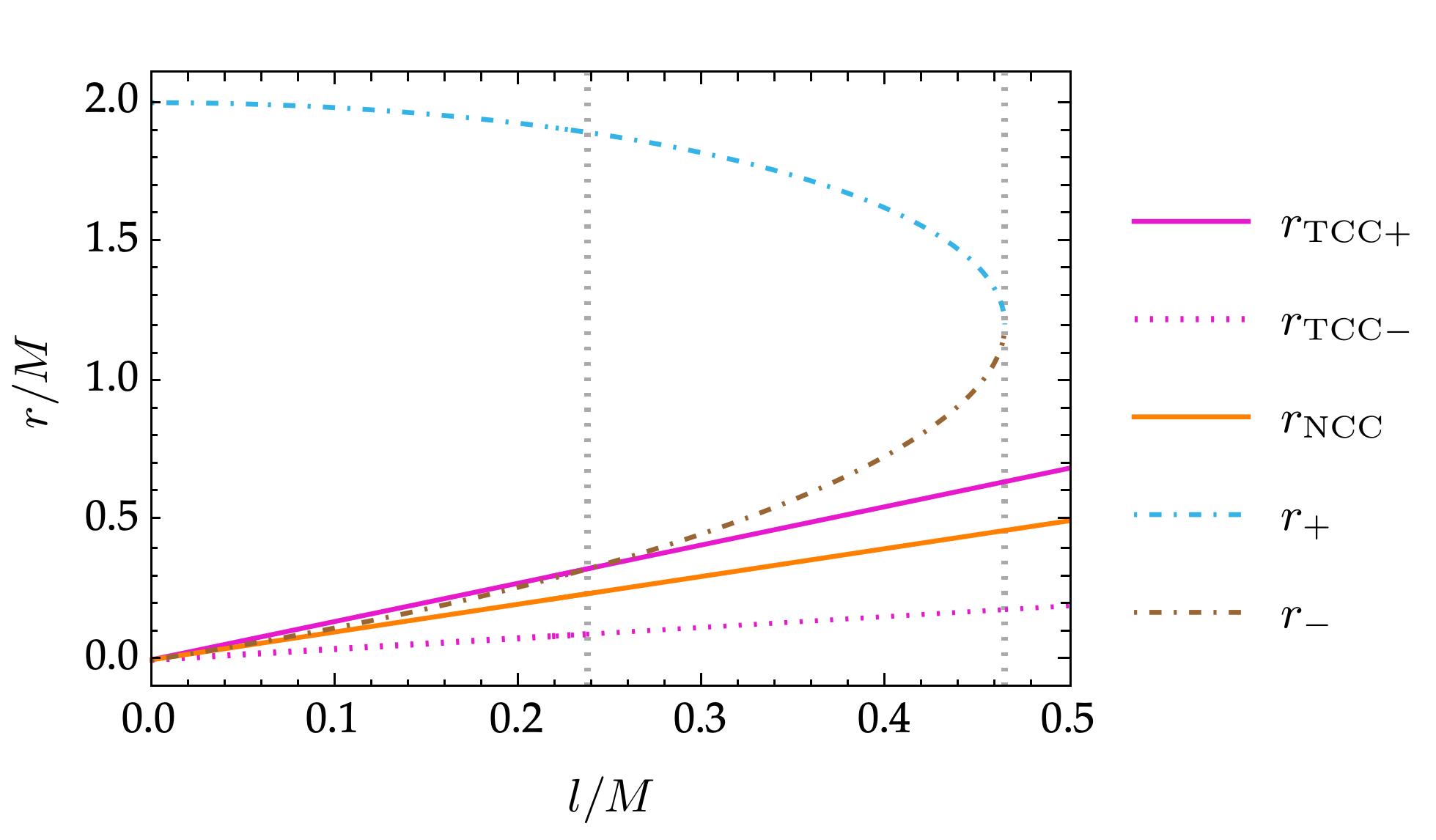}
	\caption{\label{Fig:BardeenAdSrTCCrh} Inner and outer horizons $r_\pm$, defined by $f(r_\pm)=0$, outer and inner boundaries $r_{\text{TCC}\pm}$ of the region which violates the first TCC condition, defined by $m''(r_{\text{TCC}\pm})=0$, and boundary $r_{\text{NCC}}$ of the NCC violating region, defined by $2m'(r_{\text{NCC}}) - r_{\text{NCC}} m''(r_{\text{NCC}}) =0$, as functions of the regularization parameter $l/M$ for an adS Bardeen black hole. For small enough $l/M \in (0, 0.239)$, the boundary $r_{\text{TCC}+}$ lies within the trapped region $(r_-,r_+)$ and coincides with the inner horizon  when $l/M \approx 0.465$. In turn, $r_{\text{TCC}+}$ lies in the {locally} untrapped region {inside the inner horizon} for $l/M \in (0.239,0.465 )$,  where $r_\pm/M  \approx 0.465$ marks the extremal configuration in which the two horizons degenerate to one. For larger values of $l/M$, the geometry does not describe black hole. The boundaries $r_{\text{TCC}-}$ and $r_{\text{NCC}}$ for a black-hole configuration lie always in the {locally} untrapped region surrounding the core.}
\end{figure}

More generally, the roots  $r_\pm$ of $f(r)$, representing the locations of the inner and outer horizons for the modified Bardeen spacetime with mass function~\eqref{eq:MassFunctionBardeenAdS}, can be found analytically as functions of the regularization parameter $l$. Fig.~\ref{Fig:BardeenAdSrTCCrh} shows $r_\pm$ and the outer and inner boundaries $r_{\text{TCC}\pm}$ of the region which violates the first TCC condition, where $m''(r_{\text{TCC}\pm})=0$, and the boundary of the NCC violating region, defined by $2m'(r_{\text{NCC}}) - r_{\text{NCC}}\, m''(r_{\text{NCC}}) =0$, as functions of $l/M$. For small enough $l/M$, $r_{\text{TCC+}}$ lies within the trapped region $(r_-,r_+)$, coincides with the inner horizon  when $l/M \approx 0.239$ and lies in the {locally} untrapped region {inside the inner horizon} for $l/M \in (0.239,0.465 )$, where $r_\pm/M \approx 0.465$ marks the extremal configuration in which the two horizons degenerate to one. For larger values of $l/M$ the geometry does not describe a black hole, but instead a horizonless compact object. The boundaries $r_{\text{TCC}-}$ and $r_{\text{NCC}}$ for a black-hole configuration lie always in the locally untrapped region surrounding the core.

\newpage

\section{Discussion}\label{Sec:Discussion}

The Hawking--Penrose (1970) singularity theorem~\cite{Hawking:1970zqf} weakens the causality assumption of global hyperbolicity used in the Penrose (1965) singularity theorem~\cite{Penrose:1964wq}, at the expense of stronger convergence conditions encoded in the TCC rather than the NCC. As a result, this theorem is of greater relevance (compared to the Penrose theorem) in the presence of inner horizons. By analyzing the TCC for spherically symmetric spacetimes we have established how different types of singular and regular black holes are (in)compatible with the key assumption of the Hawking--Penrose theorem. Our results allow us to identify where this theorem needs to be revised on the pathway towards a singularity theorem applicable in the presence of inner horizons.

Concretely, we have shown for a class of dynamical spherically symmetric spacetimes that the TCC decomposes into three independent conditions on the mass function, whereby only two are non-trivial in stationary spacetimes. One of these conditions is already implied by the NCC, whereas the other condition newly arises from the TCC. 
We have shown that, even though standard stationary regular black holes with a dS core satisfy the NCC globally~\cite{Borissova:2025msp}, they (locally near the core) violate the other condition implied by the TCC. Thereby these classes of regular black holes evade the theorem. On the other hand, we have analysed an explicit construction prescription for regular black holes with an adS core, starting from a regular black hole with a dS core. The resulting stationary regular black-hole spacetimes violate  the NCC near the core, and additionally violate the other condition imposed by the TCC in a finite region away from the core. Notably, the onset of TCC violation as one moves  outwards along the radial direction can occur inside the inner Cauchy horizon and therefore in a {locally} untrapped region. 

This key result suggests that understanding the nature of the timelike singularity inside an inner horizon requires an understanding under which conditions matter can focus even when light does not, i.e., in a {locally} light-untrapped region. Such conditions will be elaborated on in another paper~\cite{Borissova:2025xxx}. We expect these future directions to lay the ground for the development of singularity theorems in the presence of inner horizons.

\begin{acknowledgments}
	JB is supported by a doctoral scholarship by the German Academic Scholarship Foundation (Studienstiftung). Research at Perimeter Institute is supported in part by the Government of Canada through the Department of Innovation, Science and Economic Development Canada, and by the Province of Ontario through the Ministry of Colleges and Universities.\\
    During early phases of this work MV was supported by a Victoria University travel grant, and by SISSA, IFPU, and INFN.\\
\end{acknowledgments}

\enlargethispage{20pt}
\bibliographystyle{jhep}
\bibliography{referencesTCC}

\end{document}